\documentclass{edm_article}

\newcommand{\commentout}[1]{} 
\usepackage{graphicx,color,url}
\usepackage[super]{nth}
\usepackage{array}
\usepackage{natbib}
\usepackage[linesnumbered,ruled,vlined]{algorithm2e}
\usepackage[export]{adjustbox}
\usepackage{authblk}

\begin{document}

\title{In the Eye of the Beholder? Detecting Creativity in Visual Programming Environments}

\numberofauthors{3}
\author[1]{Anastasia Kovalkov}
\author[1]{Avi Segal}
\author[1]{Kobi Gal}
\affil[1]{Department of Software and Information Systems Engineering, Ben-Gurion University, Israel}
\maketitle

\begin{abstract}
Visual programming environments are increasingly part of  the curriculum in schools. Their potential for promoting creative thinking of students is an important factor in their adoption. However, there does not exist a standard approach for detecting creativity in students' programming behavior, and analyzing programs manually requires human expertise and is time consuming. This work provides a computational tool for measuring creativity in visual programming that combines theory from the literature with data mining approaches. It adapts the classical dimensions of creative processes to our setting, as well as considering new aspects such as visual elements of the projects. We apply this approach to the Scratch programming environment, measuring the creativity score of hundreds of projects. We show that  current metrics of computational thinking in Scratch fail to capture important aspects of creativity, such as the visual artifacts of projects. Interviews conducted with Scratch teachers validate our approach. 
%Our approach correlates with expert Scratch instructors' creativity assessment as measured in an initial trial over a group of Scratch projects. 

\end{abstract}

\keywords{} % NOT required for Proceedings
Creativity, Computational Thinking, Creativity Tests, Visual Programming Environments
\section{Introduction}

%Creativity is considered to be one of the essential tools to succeed in the 21st century~\cite{saavedra2012learning}.

Creativity is a dynamic process which generates ideas that are both novel and of value~\cite{cropley2000defining}. 
There is multiple evidence that exhibiting creativity in the classroom is linked to positive learning gains, increased motivation and to advancements in skill mastery~\cite{wheeler2002promoting,schacter2006much}.
% Specifically, creativity has been shown to promote students’ critical thinking and self self motivation and to advance students’ mastery of skills and concepts~\cite{rrr}\as{add reference}.
%As such, it has become an important component of school curriculum~\cite{wheeler2002promoting, dede2010comparing}.
Despite their increasing prevalence in schools, technological educational environments do not currently promote creativity in students’ interactions or support teachers’ ability to detect creative thinking by students.
 
 This paper provides a computational approach for detecting creativity in students' visual programming. 
 We adapt the seminal work of Torrance~\cite{torrance1965scientific} who focuses on one aspect of creativity, that of divergent thinking from conventional norms, using four dimensions: fluency (the total number of relevant ideas generated); flexibility (the number of different categories of ideas); originality (the rarity of the ideas generated); and elaboration (the amount of detail in the ideas). 
 %proposed four dimensions by which creative outcomes should be judged. (1) Originality: number of statistically infrequent solutions, (2) Elaboration: number of different ideas used in working out the details of a solution, (3) Flexibility: number of different categories of relevant solutions, (4) Fluency: number of interpretable, meaningful, and relevant solutions generated.
 We use Torrance's theory to develop a computational tool for measuring creativity in visual programming. The tool receives as input a set of visual programming projects and returns a creativity score for each of the projects in the set. 
 %We develop a separate score for each of the three Torrance dimensions of originality, flexibility and elaboration and then combine them to an overall score. 
Our tool measures divergent thinking in students' interactions from many different aspects, including programming skills and visual dimensions of the project. We combine data-driven approaches (neural nets and clustering) with Torrence's theory to measure these aspects.

We apply our approach in Scratch, a block-based visual programming language and online community targeted primarily at children. Users of Scratch can create online projects using a block-like interface. The service is developed by the MIT Media Lab, has been translated into more than 70 languages, and is used in most parts of the world\footnote{https://scratch.mit.edu/statistics}. Scratch is taught and used in after-school centers, schools, and colleges, as well as other public knowledge institutions. 

We measure creativity in two different Scratch project collections (studios) and compare the computed creativity score to existing metrics of computational thinking (CT) in Scratch. We show that while the existing CT metric capture some aspects of creativity, they fail to capture other important aspects captured by our tool that do not depend on programming skills. Supplying teachers in such schools and institutions with our tool can have a positive impact on advancing this essential capability for the 21st century.

% This paper is organized as follows. Section~\ref{related} covers related work. In section~\ref{scratchenv} we describe the Scratch programming environment. Section~\ref{interview} summarizes our interviews with multiple teachers using Scratch in the classroom. In section~\ref{compute} we describe the development of the proposed computation tool for measuring creativity in Scratch, In section~\ref{run} we run this tool on two large datasets of Scratch projects and compare it to another computational metric which does not focus on creativity. Finally, we conclude the paper with section~\ref{conclude} by discussing our results and pointing to further possible research.

\section{Related Work}
\label{related}

% Boden et. al\cite{boden1998creativity} described creativity as divided into two categories: H-creativity when the creative product is novel compared to the rest of the population and P-creativity when it comes to a leap of creativity for the individual. In addition, Cropley~\cite{cropley2000defining} also indicates identical dimensions with specifying originality measurements as within the group being tested.
Our approach builds on prior work in modeling computational thinking and creativity in visual programming environments. 

Computational Thinking (CT) expresses problems and their solutions in terms of abstraction, decomposition, and algorithmic thinking~\cite{romero2017computational}. 
There is an increasing awareness that computational thinking skills can improve decision making in everyday life~\cite{wing2006computational}, and programming classes are increasingly becoming part of the curriculum in many schools. Lye and Koh~\cite{lye2014review} review the use of several studies that use programming in K-12 classrooms to improve computational thinking skills, most of the studies reported positive outcomes.
%\as{and? what did they find}
Past work has developed tools to analyze Computational Thinking (CT) in Scratch. Dr. Scratch~\footnote{http://www.drscratch.org} is an open-source web toolkit that measures computational thinking within Scratch projects. It examines the following seven components within each project: abstraction and problem decomposition, logical thinking, synchronization, parallelism, algorithmic notions of flow control, user interactivity and data representation~\cite{moreno2015dr}. 

 Hershkovitz studied the relationship between creativity and computational thinking in a block-based multi level game framework for children's programming~\cite{hershkovitz2019creativity}. They measured computational thinking in terms of game performance. They  showed that demonstrating originality when designing the early stages of the game is associated with succeeding in this stage relatively easily, however negatively associated with progressing farther in the game.

%  The authors used two types of creativity/ First, Creative Thinking measured by using drawing task capturing 3 of Torrance's dimensions of creativity - fluency, flexibility and originality. Second, Computational Creativity was measured addressing originality dimension, it was defined by the rareness of the solution in each level of the game. In their work CT was also based on the environment variables such as maximum level reached and average number of solution attempts.
% and Computational Creativity. Creative Thinking was 
%- Torrance Test for Creative Thinking 
% Computational Creativity was measured in a block based environment presents the user with 15 mazes in which gaudiness to the destination is done by a block-based code the user is editing.
% Computational Creativity was measured for each maze (level) by one of Torrance's dimensions - originality that was defined as the rareness of the solution in the tested group.
%Computational Creativity was measured using originality in a multi level block based game environment 
% In measuring the association between these three metrics they did not find a significant relationship between Creative Thinking and CT but there was a certain relationship between computational creativity and CT in some of the levels. Demonstrating originality at a certain level reduced the number of attempts to pass it.~\cite{hershkovitz2019creativity}\as{this is too long}

Romero et al.~\cite{romero2017computational} developed a model to assess CT in students' Scratch programming and identified modeling, problem identification, code literacy and digital creativity capabilities. In their work the authors noted the existence of creative concepts in participants' projects such as using untaught and relevant Scratch elements. The projects assessment was done by experts and by Dr. Scratch. The researchers compared the differences between the scores given by the experts and the evaluation done with Dr. Scratch.
In their findings they emphasized the importance of developing automatic tools for measuring creativity in Scratch projects similarly to the evaluation done with Dr. Scratch for computational thinking.

We extend these works in several ways. First, by considering an open-ended programming environment in which students can program a wide range of different projects. Second, by including additional dimensions of creativity from the literature, including visual flexibility and elaboration. Third, by showing the generalizability of our results to hundreds of projects belonging to two different project groups and themes. Forth, we propose an automatic tool to achieve creativity assessment. 

 \section{The Scratch Visual Programming Environment}
 \label{scratchenv}
Scratch is an online block-based multi-language programming environment designed for children from elementary school to university. The environment allows users to create interactive stories, games, and animations with a focus of creating an interactive, accessible environment for all~\footnote{https://scratch.mit.edu/about}. 

Scratch blocks are shaped to fit together in ways that make syntactic sense and the environment enables the use of external data by importing photos, music clips, recorded voices and users' own graphics \cite{resnick2009scratch}. Sharing projects with the community and learning from other projects is easy and intuitive, creating a thriving and growing community of users from all over the world. 
As of today, more than 52 million people have shared more than 50 million projects in Scratch. 
Each project in Scratch is encoded into one JSON file describing the Scratch project and several additional files including the media components used in the project (e.g., sounds, images etc.). These files can be freely downloaded from the Scratch environment. 

\begin{figure}
  \includegraphics[height=1.75in, width=2.75in]{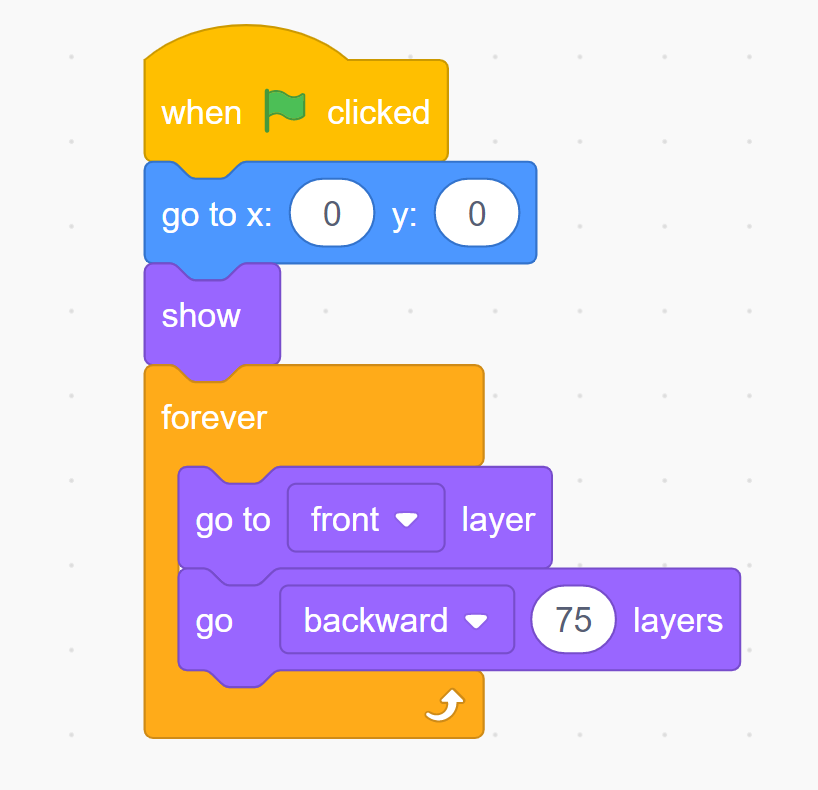}
\caption{Script in Scratch}
\label{fig:script}
\end{figure}

\subsection{Scratch Studios}

Many Scratch projects are created in the context of a Scratch Studio: an area where users share projects with the same topic or theme (e.g., Mazes, Breakout games, Explore the jungle, etc.). There are currently more than 25 million studios created by Scratch users.
%~\footnote{https://scratch.mit.edu/wiki} ~\footnote{https://scratch.mit.edu/statistics}.

%The data-sets for this work include two Scratch Studios.
%In using Scratch studios for this paper we seek to simulate the work process of a teacher in a real classroom. In such classroom scenario, the a teacher gives a shared task to the class students and asses each student work in the context of the overall class deliverables. 
For this study we specifically chose two studios that are open ended and assign minimal constraints on the programmer: Maze Games (206 projects) and Code-Your-Hero (205 projects).
The Maze Games Studio was created by users in Scratch (seventh and eighth graders). Users in this studio were asked to develop a maze based game, including a collection of paths. The goal of the developed game is to find a path from an entrance to an end goal. 

The Code-Your-Hero studio was opened as part of the Hour-of-Code activities~\footnote{https://hourofcode.com}. Users in this studio were required to produce activities (games or stories) starring their superhero. Users were given a base project that they could choose to extend, as well as a list of tutorials. 
%Unlike the Maze Game studio, the participants in the Code-Your-Hero studio were required to design a story, including characters, their mission, with many textual characteristics that depict the character of their hero and mission. See figure~\ref{fig:textual}. 

As a running example, we present a project 
from the Maze studio called The Room (Project ID 109884920). In this project, players need to move between different rooms, collect objects and activate them in order to reveal new paths towards the goal (e.g., turn on an engine to open a door). 

%(\url{https://scratch.mit.edu/projects/109884920/}). 
\subsection{Scratch Elements}
Scratch projects are created by composing elements that can broadly be divided into the following seven categories: \emph{Blocks} are pieces of code such as \texttt{GoTo}, \texttt{MouseDown}, \texttt{WaitUntil}. \emph{Costumes} contain elements for manipulating characters and backdrops 
%elements for manipulation of elements such as characters and backdrops,
and are commonly used for animations. \emph{Sounds} contain audible elements that are pre-included (e.g., Pop, Meow) or designed by the user. \emph{Monitors} include values of variables or lists. 
\emph{Arguments} accept user input, such as Message, Color, Direction etc. 
\emph{Action keys} trigger actions in the program, and contain the English alphabet, the number keys, the arrow keys and the space key. \emph{Extensions}, refer to external hardware or blocks allowing for more advanced functionality in the program including the ones created by the user (e.g., Video, Text to Speech etc.). 
In all, the size of the possible set of elements across all categories is 413 in the Code-Your-Hero projects and 652 in the Maze projects.

A \emph{script} in Scratch is a collection of ordered blocks that generate a program logic. 
Figure~\ref{fig:script} shows one of the scripts of The Room project.
It contains elements from the Block category (\texttt{WhenGreenFlagClicked}), the Argument category (a Number variable with value ``75" and X and Y variables with value ``0" ). 
%On top of this categories and features we are looking into the visual and textual elements of the program.
%Additionally we address low level features in our measurements including combinations of blocks that produce logical behavior in the program referred as Script and their max depth ($Md$). The presented Script in 

\begin{figure}
  \includegraphics[height=1.5in, width=3.25in]{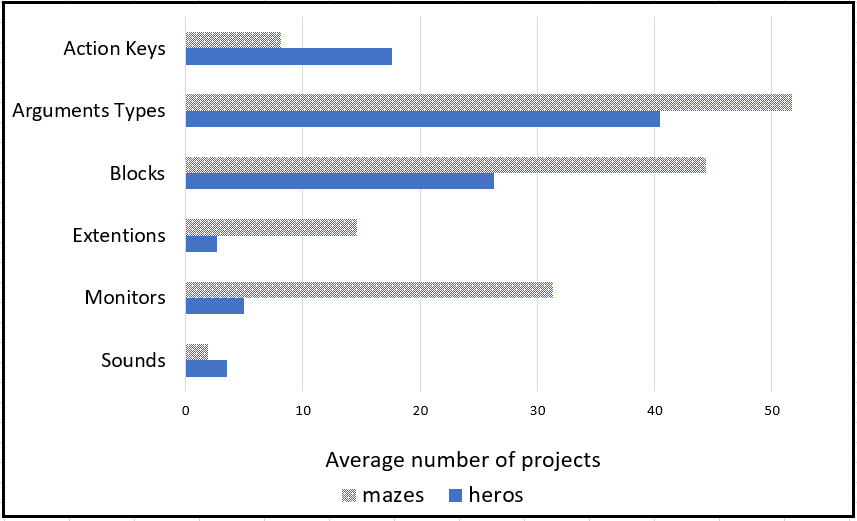}
\caption {Histogram of categories in the two studios}
%. Originality can be expressed by using elements from categories with low frequency in the projects (Keyboard keys and Sounds) than from common categories such as blocks and input types (Condition, Message, String etc.)}
\label{fig:frequency}
\end{figure}

Scratch projects vary widely in the use of elements from each category. To illustrate, 
Figure~\ref{fig:frequency} shows a histogram of the major categories in the projects 
of our two studios, shown in the $y$ axis. The $x$ axis lists the number of projects that contain elements of each category, averaged over the number of elements in the category. As shown by the figure, in both studios, arguments and blocks categories are widely used, while sounds and extensions are much more rare. Many of the elements in the sound and extension categories are designed by the users themselves, rather than prepackaged by Scratch. 

% It can be concluded that due the uncommon use of elements that belong to the Sound, Extensions and Keyboard keys programs that will use such elements (imported sounds, the text to speech extensions or infrequent keys like numbers) will be considered as more original in it elements.

% ** move later **
% In the Code-Your-Hero Studio the project considered as most creative by the proposed model called My Hero: MILITARY MOVES (Project ID 352121127). Its a game in which players are required to overcome boredom by activities like collecting balls and throwing sparks. The game includes many elements from the non-common categories in this Studio like 3 types of Extensions including Music, Pen and Text to Speech as well as a monitor that shows the boredom we collected, and varies type of keyboard such as a,w,d,s for movement and space for throwing.

%The Monitors category was uncommonly used in the Code-Your-Hero Studio but we observe that the three of the five considered most original in this Studio by our measurements are using Monitors in their project. One of them 
 %add heroes exp 

 % The Room project, considered as the most original project in the Maze Studio uses 123 of these elements (Space keyboard, Sensing Mouse Y block etc). 
 % ~\footnote that was 
%special in several ways. 

\section{Interviewing Scratch Educators}
\label{interview}
To better understand how Scratch educators are considering creativity in their students' work, we have conducted interviews with three Scratch teachers working with students between the ages of 9 to 14. Two of the teachers work with Scratch as part of the  curriculum in school, and the third teacher uses Scratch in after school computer science lessons.
%The experts agreed that they were mainly focused on the algorithmic and programming part of the environment, the lesson plans are more focused on understanding loops, program planning and creating scripts depending on the task they were given. But, 
All teachers identified creativity as an important element in their Scratch teachings. They described creativity as an ability they look at, support and encourage in addition to focusing on computational thinking skills. This stresses the need for a creativity focused metric to assist teachers in their work. 
%elements of their students while working with Scratch.

The teachers emphasized the importance of novelty with respect to students' project, whether in context to the history of a specific student, or the class.
%This input emphasizes the need to measure originality as one aspect of creativity, and to do this in context - in relation to other projects. 
One teacher also mentioned the importance of measuring creativity as an evolving aspect for each student in separation, i.e. comparing a student's project to earlier projects they did.
%This is a promising direction and we reserve it for future work. 

Teachers view  creativity in Scratch as being expressed in two manners: (1) in the Scratch code itself - the blocks and other Scratch elements and (2) in the Scratch project output as expressed in the visual and textual artifacts presented when the project is executed. Teachers stressed the importance of creativity in the design process. Projects are considered more creative if they diverted from baseline projects presented in the classroom. Additionally, two teachers described wit and humor in the project outcome (specifically in the text messages and images presented) as creative elements.
%This feedback emphasizes the need of any computed creativity metric to look both on the Scratch project code as well as on the Scratch project artifacts such as visual graphics and textual messages. 
% Specifically, in Torrance's creativity theory this information may be captured by the flexibility dimension.

All teachers also emphasized the plurality of scripts, characters and blocks used as an indication of the creative capabilities of projects created by students. This connects with the Torrance's elaboration dimension which considers the quantity of the elements created. 
 
All teachers expressed the need to combine a computational tool for creativity with  the personal connection and familiarity of an educator with the students and their learning process. The teachers emphasized the need to view any such score as an auxiliary tool that may assist teachers, always leaving decision making and final call in the human hands of the teacher. 

\section{Modeling Creativity in Scratch}
%\section{Combined Creativity Score (CCS)}
\label{compute} 
We now describe the development of a computational tool for measuring creativity in Scratch. Our goal was to provide computational measures of the dimensions given by Torrance in Scratch and to compute a creativity score for any project. 

%We seek to measure multiple aspects of a Scratch project that relate to the various Torrance dimensions. As noted earlier, these dimensions are judged in context, i.e., in relation to other projects, Thus, our metrics are developed in the context of a Scratch Studio, as explained below. 
%To support the assessment of students’ creativity in Scratch, we focus on developing a computational
%metric for creativity, 
The tool receives as input a set of Scratch projects from a given studio and returns a creativity score for each of the projects in the set. The set of projects can belong to the same user or to different users. 
%This metric should have high resemblance to the manner by which education and creativity experts evaluate students’ creative outcomes. 
The metric should support the analysis of creativity in a wide range of Scratch projects across multiple project topics and themes. 
We describe a separate score for each of the three dimensions originality, flexibility and elaboration. As our metrics score is given to each project individually and not to a user we will not refer to the fluency dimension as it refers to the quantity of user-delivered products.
% We did not refer the fluency (quantity of ideas) dimension due to our inability to control the number of projects each user uploads into a studio.\as{not clear} 
%We compare our model's ability to capture the creative aspects of a Scratch project with Dr. Scratch, a tool that is focused on measuring computational thinking of Scratch programs. 

% We now give a brief description of the data-sets used in this work before describing the computational metrics developed. 

%The project JSON file includes all the information about the sprites, scripts, monitors, extensions and the media used in the project. Thus, parsing and analyzing this file is the first step in a Scratch program evaluation.
%We now move to describe the creativity m 
%To this end,
%We adapt the dimensions. The and calculate the Combined Creativity Score (CCS). We did not refer the fluency metric due to our inability to control the number of projects each user uploads into a studio. \kg{what does this mean?}

\begin{figure}
    \includegraphics[height=1.5in, width=3.25in]{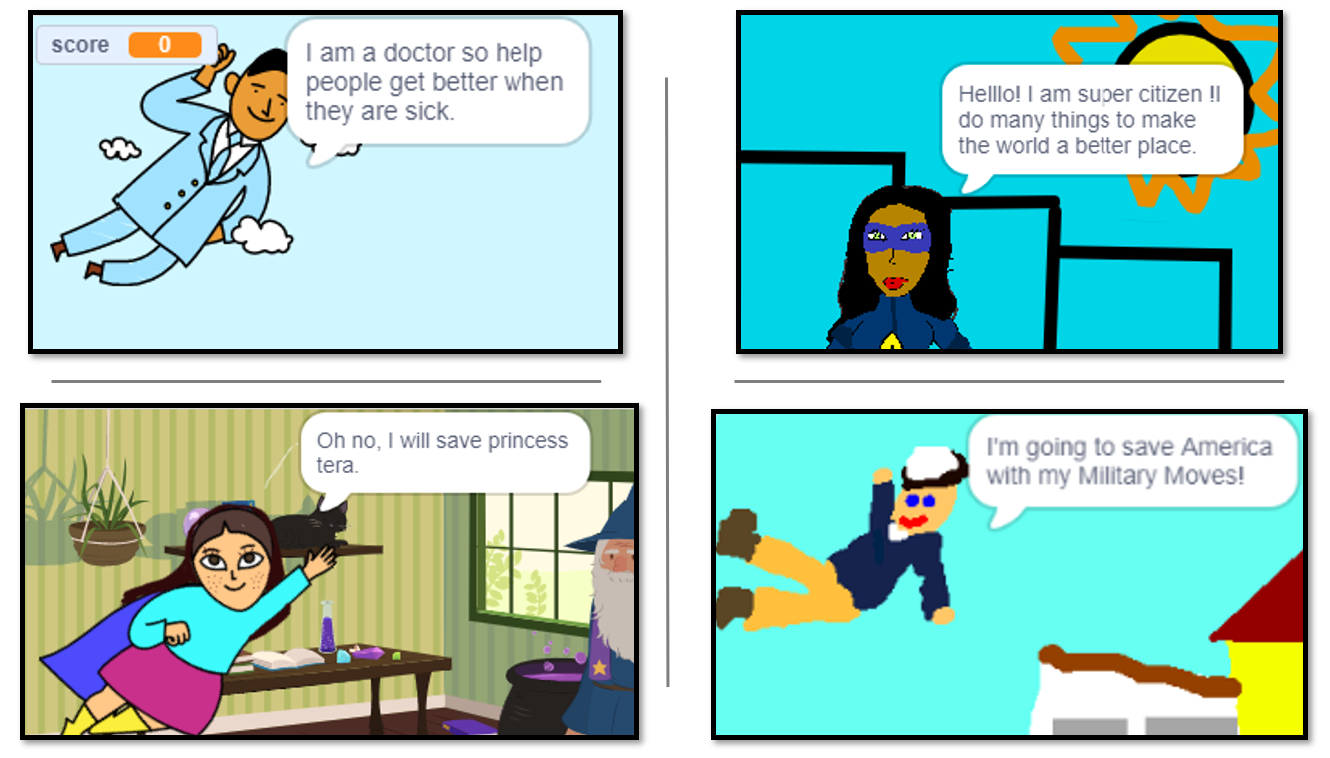}
\caption{Examples of text elements in Code-Your-Hero projects}
\label{fig:textual}
\end{figure}
%\subsection{Creativity Measurements}

% \textbf{Combined Creativity Score (CCS)}.

%The metrics are computed in the context of a Scratch Studio, a collection  of projects with the same theme.
The \emph{originality} score of a Scratch project $j$ is measured with relation to all projects in the studio ($D$). A Scratch project is more original if it uses more elements that are rarely used by other projects in the studio. 

% When discussing the originality of a project we consider six of the elements categories (Blocks, Extensions, Monitors, Sounds, Input types, and Keyboard keys). \kg{the following can come after the definition of elements. Where is your running example?}

Let $El_j$ be the set of elements in project $j$. 
For each element $e_i$, let $\#(e_i)$ denote the number of studio projects that this element appears in. The uniqueness of each element is $O(e_i)= 1/ \#(e_i)$. For example, the Block element \texttt{WhenGreenFlagClicked} that is used in all maze projects gains a low uniqueness score of $0.004$ while the Action key element \texttt{m} (with the purpose of showing the full map of the game) that is used only in 3 maze projects gains a higher uniqueness score of $0.333$. 

The originality of the project $j$ is the sum of the uniqueness scores of all its elements as given by 
  $O_j= \sum_{e_i\in El_j} O(e_i)/Z$, where $Z$ is a normalization factor over all of the projects in the studio.
For example, the project The Room uses 51 elements that are used in less than 15\% of the projects, 
%and only 9 elements used in all of them which
leading to the highest originality score in the Maze studio.
% Let $E_j$ be the set of elements in project $j$. The possible set of elements \kg{I didn't understand difference between elements and categories}  block categories (Motion, Event etc.), \kg{goto, motion    is an example, how many elements in total? can you list them all?}block types (goto, say), sounds, costumes, types of blocks inputs (Condition, Measurement etc.), extensions and monitors.
% For each element  $e_i$, let $n(e_i,D$) denote  the number of studio projects that this element appears in. The uniqueness of each element is $O(e_i)= 1/ n(e_i,D)$. 
% \kg{explain why this makes sense, use the running example. Use $n(e_i)$ for describing the number of something throughout the per, e.g., the number of projects containing $e_i$}

% % originality measurements
% \begin{equation}
%     O(e_i)= {number\>of\>projects\>containing\>element\>i} \label{eq:usage} 
% \end{equation}

The \emph{elaboration} score of a Scratch project $j$ is based on 
 the number of occurrences of elements in the different categories as well as the structure of the scripts making up the project. 
%Intuitively, a project with more elements is a more elaborated project.
First, we count how often elements from each of the seven categories appears in a project (Blocks $B$, Costumes $C$, Sounds $S$, Action keys $K$, Extensions $Ext$, Monitors $M$ and Arguments $A$).
%Specifically, we summarize for each element the number of its appearances in the project.
To illustrate, the room project uses 65 unique elements from the block category but many of them are used more than once, so overall it contains 2742 block elements.

Second, we count 
the amount of scripts ($Sc$) in the project and their max depth ($Md$). The elaboration score of a project is given by 
$ E_j = \big((\#(B_{j}) + \#(Sc_{j}) + \#(K_{j}) + \#(A_{j}) 
   \\ + \#(Ext_{j}) + \#(M_{j}) +\#(C_{j}) + \#(S_{j}) +\#(Md_{j})\big) /Z$, where $Z$ is a normalization factor.
% $E_j$ is given in Equation~\ref{eq:elaboration}. 
% elaboration measurements
% \begin{equation}
% \label{eq:elaboration}
% \begin{split} 
%     E_j =   & \big( \ (\#(B_{j}) + \#(Sc_{j}) + \#(K_{j}) + \#(A_{j})  
%      \\
%      &  +  \#(Ext_{j}) + \#(M_{j}) +\#(C_{j})  + \#(S_{j}) +\#(Md_{j})\  \big) /Z
% \end{split}
% \end{equation}

% In our example The Room project contains elements from each of the seven categories (Blocks $B$, Costumes $C$, Sounds $S$, Action keys $K$, Extensions $Ext$, Monitors $M$ and Arguments $A$). 
% including 65 types of blocks that contain 2742 elements ,2 extensions, 3 types of monitors on 41 variables (changeable values recorded in Scratch's memory), 32 sounds, 10 types of arguments with over 3500 values ($Av$) and 1 type of keyboard keys. It also contains almost 300 costumes and 90 scripts with max depth ($Md$) of 42. Over all it's ranked \nth{4} in the Maze studio by this metric.
To illustrate, The Room project is built from 90 scripts with max depth of 42, Over all it has an elaboration score of $0.637$ and is ranked \nth{4} in the Maze studio by this metric.
% To illustrate, The Room project contains elements from each of the seven categories. These include 65 blocks elements appearing over 2500 times in the project combined into 90 scripts with max depth of 42, 2 extensions, 3 monitors duplicated on over 40 variables (changeable values recorded in Scratch's memory), The project contains 32 sounds 299 costumes, it uses over 3500 arguments and one action key. Over all it has an elaboration score of $7066$ and is ranked ranked \nth{4} in the Maze studio by this score.
% \kg{you need to say it's elaboration score. Also it looks like the elaboration score will always be much higher than the originality score, so how do you handle this issue?}\ak{when we combine all metrics in line 268 we describe the normalization of each}

% Where, \kg{how is $S(P_j)$ computed, again you need an example. Is it parallel to $O(P_j)$}
% $
% \mathbf S_{p_j}- sounds, C_{p_j}- costumes, E_{p_j}- extensions,\\
%  \mathbf M_{p_j}- monitor, I_{p_j}- inputs and K_{p_j}- key \ board \ interactions 
% $
% are the Scratch's elements considered. 
 The \emph{flexibility} score of a Scratch project is based on the diversity that is embedded in the textual and visual outputs of the project. Figure~\ref{fig:textual} presents this diversity with examples from the Code-Your-Hero studio. While all outcomes demonstrate the adherence to the requirements of presenting their hero. In each project presented, characters and backgrounds differ, as well as various textual outputs depicting the hero's role. 
 %the amount of textual and visual categories (when "categories" mean different "types" of textual or visual solutions). in the project output. Intuitively, a project which produces more "categories" of textual or visual elements is considered  a more flexible project  \kg{need to define textual and visual categories}.

\begin{figure}
\includegraphics[height=2.25in, width=3.5in]{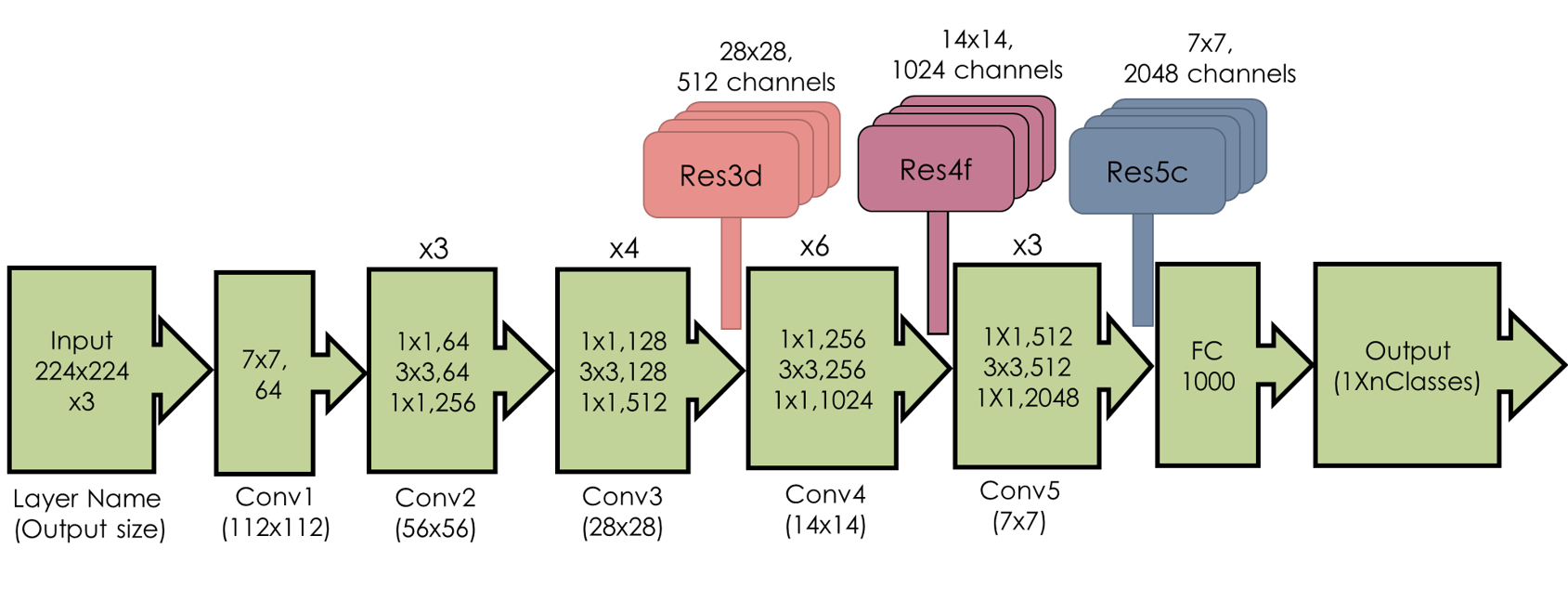}
\caption{ResNet Neural Network}
\label{fig:resnet}
\end{figure}

We measure the diversity of the visual outputs by clustering  the images of all projects in the studio  into 
different groups.
 To this end we used a ResNet50 convolutional neural network~\cite{he2016deep} to transform each image in this collection to a 2048 vector representation. ResNet50 is a neural network trained on more than a million images from the ImageNet database~\cite{imagenet_cvpr09}. The network is 50 layers deep and can classify images into 1000 object categories, such as keyboard, mouse, pencil, and many animals. 
Figure~\ref{fig:resnet} presents the architecture of the ResNet50 neural network. 

To group together projects with similar image designs, we applied a K-Means clustering algorithm on the image vectors, which produces the different image clusters for the given studio. Finally, the visual flexibility score of a project $j$ was computed as the number of different clusters to which this project's images belong. Algorithm~\ref{alg:vf} summarizes this process. Line 1 embeds a set of vectors from all of the images of the projects in the studio. Line 2 applies a Means clustering algorithm on the vectors. Each image is assigned to a cluster. Lines 3-5 assign to each project $j$, a visual flexibility score that is the cardinality of the set of unique clusters that match all of its images. 

\begin{figure}
\includegraphics[height=1.75in, width=3in]{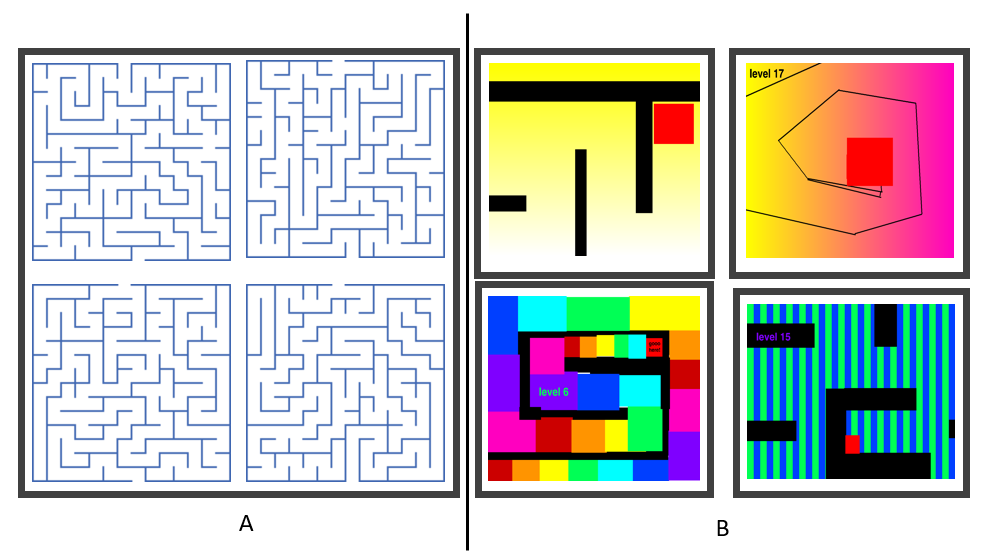}
\caption{Images from two Maze projects with different visual flexibility score.}
\label{fig:compare_projects}
\end{figure}

To illustrate, 
 Figure~\ref{fig:compare_projects} shows two projects in the Maze studio with different visual flexibility scores, as determined by the algorithm. Project A (left) and Project B (right) both have four maze images. The images of Project A  are visually similar to each other. They are clustered into one group and the project gets a low visual flexibility score. The images of Project B are visually different from each other (colors and graphics wise). The images are clustered into four groups and the project gets a high visual flexibility score.

\begin{algorithm}[t]
\KwData{Studio images - Images; Studio projects - Projects}
\KwResult{Visual flexibility scores for all projects in studio}
\Begin{
    $vectors \gets ResNet50(Images)$\;
    $clusters \gets KMeans(vectors)$\;
    \For {$P_{j} \in Projects$}{
        $clusterIDS_j \gets Set(clusters[Images[j]])$\;
        ${Vf}_j \gets |clusterIDS_j|$\;
        $Vf.insert(Vf_j)$\;
    }
    \KwRet {Vf} \;
}
\caption{Computing Visual Flexibility Score}
\label{alg:vf}
\end{algorithm}

% Similarly, to compute the textual categories of a Scratch Studio we collected the textual elements that appeared in the Studio projects in Argument elements such as message, broadcast question 
Similarly, we measure the diversity of the the textual outputs by clustering the textual elements that appeared in the studio projects into different groups. The textual elements based on the Argument elements such as Message, Broadcast, Question etc and they converted to vectors using TF-IDF model~\cite{ramos2003using}.

% which allows the determination of weight for each term (or word) in each document.~\cite{ramos2003using}.
% To illustrate, the Room project contains 21 textual terms that belongs to four separated clusters. Terms like ``level done" and ``level lighton" share a cluster, but the terms ``start" and ``end" belongs to another cluster.

% with 
% Each textual element (term or sentence) was represented by a vector that was calculated using a term frequency-inverse document (TF-IDF). TF-IDF is a numerical model which allows the determination of weight for each term (or word) in each document.~\cite{ramos2003using}.
% We applied k-means clustering on these vectors to identify the textual categories of a Scratch Studio. Then the textual flexibility ($Tf$) score of a project $j$ is the amount of clusters to which the project's text messages belong.
% To illustrate, the Room project contains 21 textual terms that belongs to four separated clusters. Terms like ``level done" and ``level lighton" share a cluster, but the terms ``start" and ``end" belongs to another cluster.

The flexibility score of project $j$ is  the sum of the textual ($\#Tf_{j}$) and visual ($\#Vf_{j}$) flexibility scores given by $F_j= (\#Tf_{j}+\#Vf_{j}) /Z$ where $Z$ is a normalization factor.
%as shown in Equation \ref{eq:flexibility}.
For the project The Room the over all visual and textual amount of categories is 17 resolving the highest flexibility score in the Maze studio.

% \begin{equation} \label{eq:flexibility}
%     F_j= (\#Tf_{j}+\#Vf_{j}) /Z 
% \end{equation}
% where $Z$ is a normalization factor.

Finally, the \emph{Combined Creativity Score (CCS)} is the normalized summation of the originality, elaboration and flexibility scores, as given in Equation (\ref{eq:ccs}).
\begin{equation}\label{eq:ccs} 
CCS_j = (O_j+E_j+F_j)/Z
\end{equation}
where $Z$ is a normalization factor.
The Room project received the highest score in the Maze studio.

\section{Comparing CCS to Dr. Scratch}
\label{run}
 In this section we study the relationship between CCS, our proposed creativity metric, to Dr. Scratch, which is used to measure computational thinking, over all projects in the studios. We investigate the differences in the ranking of projects created by both metrics. We will refer to the project based on their ID given by Scratch and can be accessed at  \url{https://scratch.mit.edu/projects/ID}.
 %As Dr. Scratch focuses on measuring computational thinking, we expect to see differences in the project rankings obtained by these two metrics and are interested in examining these differences. 

% To better understand the differences between the two metrics we look at the top 5 projects ranked highest by each of the metrics in both projects types. 
Table~\ref{tbl:fivetop_maze} shows the comparison between CCS and Dr. Scratch for the Maze projects and Table~\ref{tbl:fivetop_heroes} presents this comparison for the Code-Your-Hero projects. In each such table we present the top five ranked projects by each metric and the corresponding ranking of every project by the other metric. We also show an image of each project in the table.
As can be seen in both tables, projects ranked highest by one metric are not necessarily ranked highest by the other metric.

\begin{table}[t]
\centering
\caption{Top Ranked Maze Projects}
\label{tbl:fivetop_maze}
\resizebox{.47\textwidth}{!}{
\begin{tabular}{|m{0.12\textwidth}|c|c|m{0.12\textwidth}|c|c|}\hline
\multicolumn{3}{|c|}{CCS Ranking}&\multicolumn{3}{|c|}{Dr. Scratch  Ranking} \\\hline
Project Image & Rank & Dr. Scratch Rank & Project Image & Rank & CCS Rank\\\hline
\includegraphics[width=0.12\textwidth,margin=0pt 1ex 0pt 1ex,valign=m]{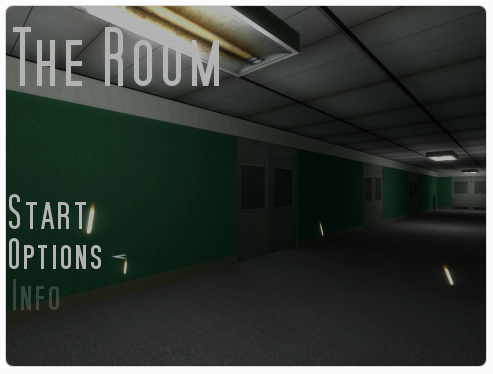}& 1 & 8 &
\includegraphics[width=0.12\textwidth,margin=0pt 1ex 0pt 1ex,valign=m]{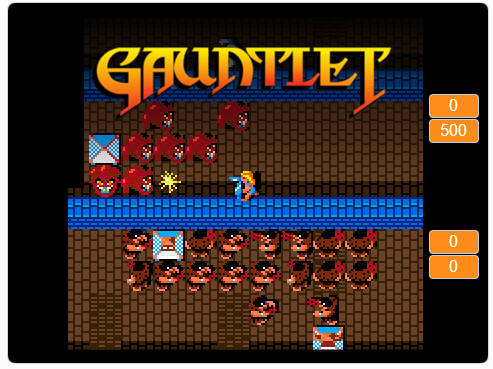}& 1 & 2 \\\hline
\includegraphics[width=0.12\textwidth,margin=0pt 1ex 0pt 1ex,valign=m]{images/Mazes/ccs_2.png}& 2 & 1 &
\includegraphics[width=0.12\textwidth,margin=0pt 1ex 0pt 1ex,valign=m]{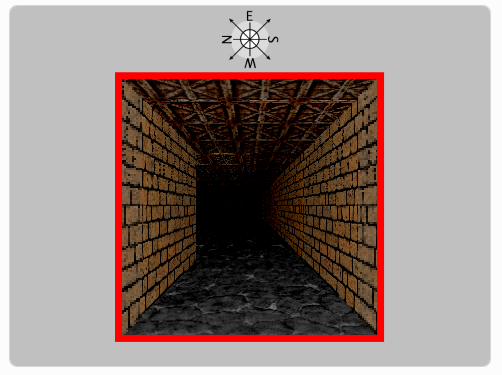}& 2 & 3 \\\hline
\includegraphics[width=0.12\textwidth,margin=0pt 1ex 0pt 1ex,valign=m]{images/Mazes/ccs_3.png}& 3 & 2 &
\includegraphics[width=0.12\textwidth,margin=0pt 1ex 0pt 1ex,valign=m]{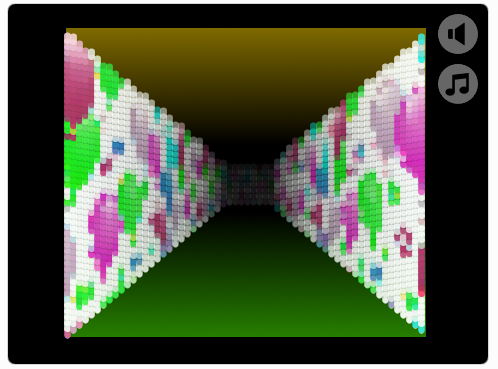}& 3 & 9 \\\hline
\includegraphics[width=0.12\textwidth,margin=0pt 1ex 0pt 1ex,valign=m]{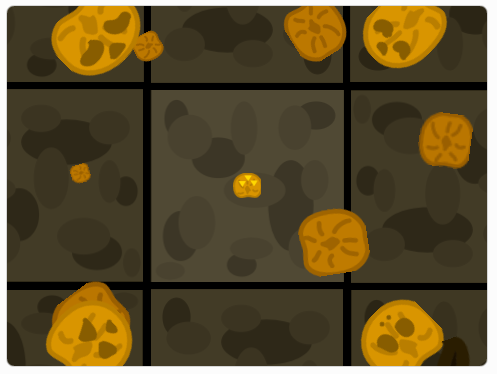}& 4 & 27 &
\includegraphics[width=0.12\textwidth,margin=0pt 1ex 0pt 1ex,valign=m]{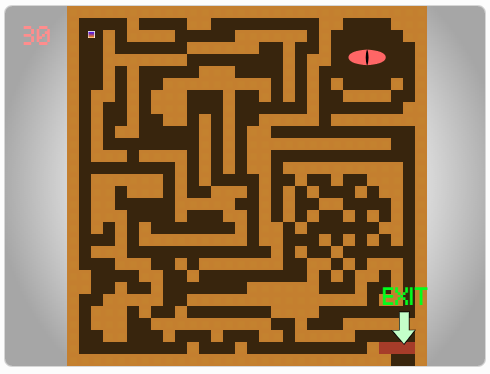}& 4 & 10 \\\hline
\includegraphics[width=0.12\textwidth,margin=0pt 1ex 0pt 1ex,valign=m]{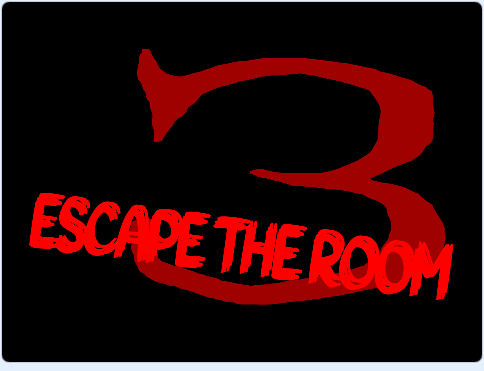}& 5 & 9 &
\includegraphics[width=0.12\textwidth,margin=0pt 1ex 0pt 1ex,valign=m]{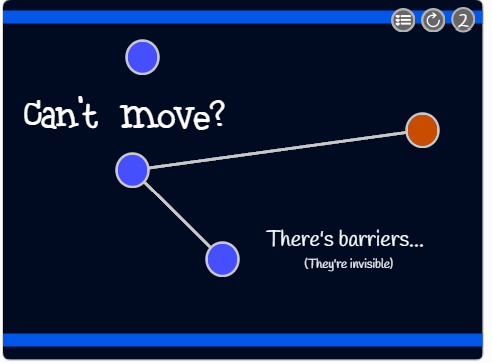}& 5 & 17 \\\hline
\end{tabular}
}
\end{table} 

\begin{table}[t]
\centering
\caption{Top Ranked Code-Your-Hero Projects}
\label{tbl:fivetop_heroes}
\resizebox{.47\textwidth}{!}{
\begin{tabular}{|m{0.12\textwidth}|c|c|m{0.12\textwidth}|c|c|}\hline
\multicolumn{3}{|c|}{CCS Top Ranking}&\multicolumn{3}{|c|}{Dr. Scratch Top Ranking} \\\hline
Project Image & Rank & Dr. Scratch Rank & Project Image & Rank & CCS Rank\\\hline
\includegraphics[width=0.12\textwidth,margin=0pt 1ex 0pt 1ex,valign=m]{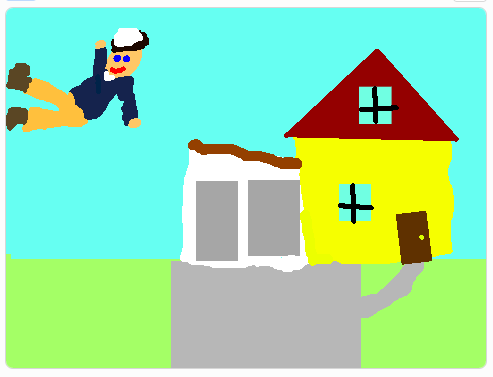}& 1 & 1 &
\includegraphics[width=0.12\textwidth,margin=0pt 1ex 0pt 1ex,valign=m]{images/Heroes/ccs_1.png}& 1 & 1 \\\hline
\includegraphics[width=0.12\textwidth,margin=0pt 1ex 0pt 1ex,valign=m]{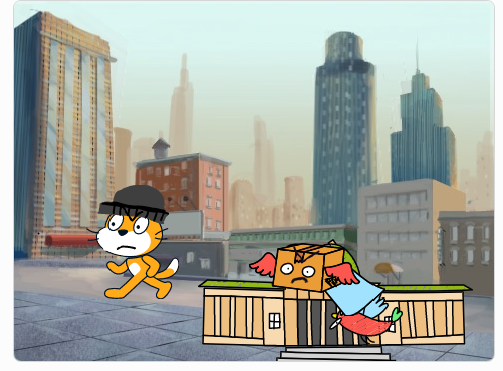}& 2 & 14 &
\includegraphics[width=0.12\textwidth,margin=0pt 1ex 0pt 1ex,valign=m]{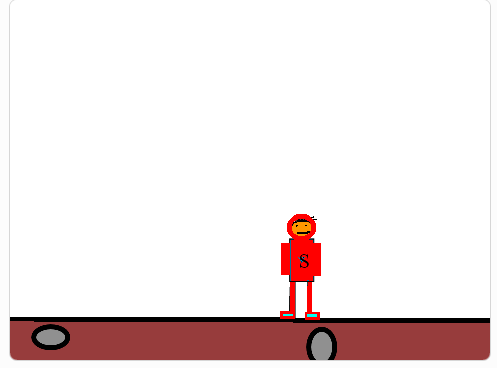}& 2 & 19 \\\hline
\includegraphics[width=0.12\textwidth,margin=0pt 1ex 0pt 1ex,valign=m]{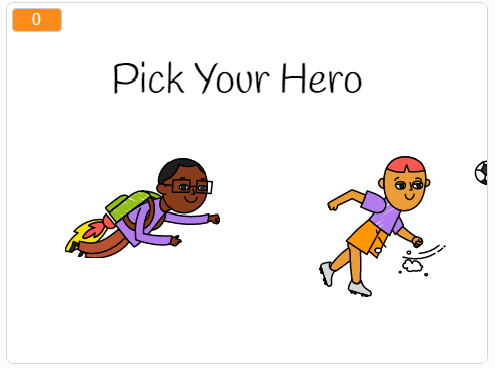}& 3 & 4 &
\includegraphics[width=0.12\textwidth,margin=0pt 1ex 0pt 1ex,valign=m]{images/Heroes/ccs_2.png}& 3 & 3 \\\hline
\includegraphics[width=0.12\textwidth,margin=0pt 1ex 0pt 1ex,valign=m]{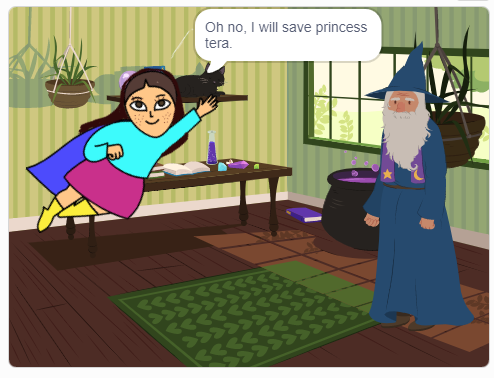}& 4 & 38 & 
\includegraphics[width=0.12\textwidth,margin=0pt 1ex 0pt 1ex,valign=m]{images/Heroes/ccs_3.png}& 4 & 5 \\\hline
\includegraphics[width=0.12\textwidth,margin=0pt 1ex 0pt 1ex,valign=m]{images/Heroes/ccs_2.png}& 5 & 3 &
\includegraphics[width=0.12\textwidth,margin=0pt 1ex 0pt 1ex,valign=m]{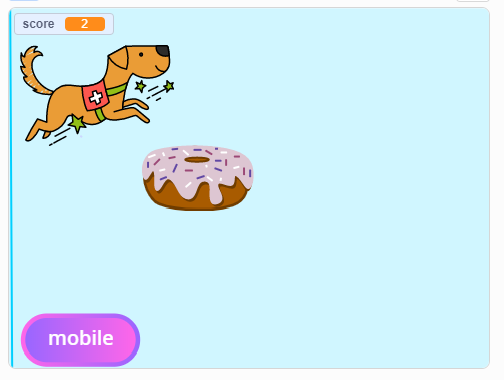}& 5 & 156\\\hline
\end{tabular}
}
\end{table}

Table~\ref{tbl:correlations} presents the Kendall Rank Correlation Coefficient~\cite{abdi2007kendall} between all the projects in the studio. 
%between two ranked lists: a CCS based project ranking list and a Dr. Scratch based project ranking list.
As shown in the table, there is some correlation between CCS and Dr. Scratch in the two project types, although the metrics do not fully agree. Specifically, we see a higher correlation between the two measures in the Maze projects ($\tau=0.628$) than in the Code-Your-Hero projects ($\tau=0.549$). Maze projects are generally more complex than Code-Your-Hero projects from a programming perspective. For  projects where creativity is demonstrated in ways other than programming skills in Scratch, such as with images, text and sounds, the correlation between CCS and Dr. Scratch  decreases.
 
\begin{table}
\centering
\caption{Ranking correlations between CCS and Dr. Scratch}
\label{tbl:correlations}
\begin{tabular}{|c|c|c|} \hline
&Maze projects&Code-Your-Hero projects\\ \hline
Kendall's $\tau$ &0.628 & 0.433 \\ \hline
p-value & $5.496e^{-41}$ & $2.789e^{-20}$\\ \hline
\end{tabular}
\end{table} 

%Although in both types of projects there is a correlation between the two measurements, as represented in~\ref{tbl:correlations} but this is not a consensus. It may be implied that when higher programming capabilities exist, more creative skills can be expressed thus we can see that in the maze projects that contained more complex and detailed project there is a higher correlation between CCS and Dr. Scratch ($\tau=0.628, \rho=0.791$). However, when observing the ranking correlation in the hero's project we see a decreasing we see a decline in similarity ($\tau=0.549, \rho=0.571$).  

%For the maze projects, we get the average of 0.190 (SD = 0.168) creativity score, an average of 0.087 (SD = 0.129) for originality score,  an average of 0.081 (SD = 0.152) for elaboration score and an average of 0.278 (SD = 0.173) for flexibility score. 

%The measurements of hero projects are a little bit higher. The average score of hero projects is 0.315 (SD = 0.161), they got an average of 0.078 (SD = 0.110) for originality score,  an average of 0.198 (SD = 0.178) for elaboration score and an average of 0.374 (SD = 0.210) for flexibility score. 
For the Maze studio, CCS ranked highest a project from our running example called The Room (ID 109884920).
%Items collected by players provided clues (e.g. ) ; 
%the project was the \nth{4} highest in terms of number of elements (over 7000 elements),
This project was the \nth{4} highest in terms of elaboration, and \nth{1} in terms of flexibility and originality. 
Clearly, this project exhibits many aspects of creative thinking, yet it was ranked only \nth{8} by Dr. Scratch. 
%Dr. Scratch places fifth a project with relatively low CCS score (0.417). In this maze user moves from point a to b by moving across a graph, but based on the metrics it can be understood, although most of the metrics are higher than average (flexibility - with 9 visual categories and elaboration with 76 costumes and 73 scripts its blocks and other elements are widely used, it has no unique elements and only 3 elements that used in less than 10 other projects, and so placed 19th by CCS.
In the same studio, CCS ranked at the \nth{4} place a project that Dr. Scratch ranked at \nth{27} place. This project (ID 339733364) is a 3D Maze game
where points can be gained by collecting coins and candies. Additionally, there are obstacles represented by other characters with special abilities such as shooting. The only way to reach the target is by collecting keys and revel new paths by using them.
The project ranked \nth{1} in terms of elaboration (122 scripts, 76 costumes, 3416 blocks). This is an original project with varied details (thus high on the originality and flexibility scores - with 12 different visual clusters and 3 textual ones) which was not captured by Dr. Scratch. 

Similarly, there is disagreement in the 5 top Code-Your-Hero projects. Although CCS and Dr. Scratch rank \nth{1} the same project, the \nth{2} project by Dr. Scratch was ranked \nth{19} by CCS. This project (ID 344287265) has low diversity in the visual flexibility outputs (4 different clusters out of the possible 17) and is missing textual elements compared to projects ranked higher by CCS in this studio (2 different clusters out of the possible 12). It ranked \nth{14} in terms of elaboration with 1040 elements while the project ranked highest under this metrics uses almost twice as much. 
The project (ID 352557648) ranked \nth{5} by Dr. Scratch but only \nth{156} according to the CCS metric. In this project the main character is a flying dog trying to catch donuts. It is ranked high by Dr. Scratch due to features such as parallelism and synchronization, but in relation to the other projects, it has minimal textual flexibility with 1 unique cluster. The visual outputs used are limited and taken from the Scratch environment. In addition, it contains no unique elements at all and most of its elements (including 71 blocks, 10 scripts, 3 sounds) are similar to many other projects in the studio. This occurs due the similarity of this game to the tutorials that has been shown in the activity. Thus, while this project is considered high on the list of Dr. Scratch, CCS gives it a relatively low score. 

%Even the amount of elements used in this project are relatively low (71 blocks, 10 scripts with a maximum depth of 11) but probably contains the ones observed by Dr. Scratch.

We note another interesting phenomena observed in the Code-Your-Hero studio. In this studio both metrics ranked two projects that were developed by the same user, one project  (ID 353026680) improving the other (ID 352632009) and implemented at a later stage. Over 100 elements were added including 1 costume and over 50 blocks combined into 10 new scripts, including a unique element in the Action keys category.
%The "improvement" project was based on the same idea as the earlier project and had a very similar implementation with some additions. 
These projects were ranked at the \nth{3} and \nth{5} places by CCS, with the later improved project ranked higher. However, Dr. Scratch was not able to detect the subtle changes between the projects and ranked them both with the same score (\nth{3} and \nth{4} for Dr. Scratch).
%When observing these two projects in detail, we notice that the change that CCS managed to capture the addition of new elements  These changes were missed by Dr. Scratch which is not targeted at measuring such changes. 
% \begin{equation}\label{eq:RSquared_eq}
%     Correl(X,Y) = \frac{\sum{(x-\bar{x})(y-\bar{y})}}{\sqrt{\sum{(x-\bar{x})^2}\sum{(y-\bar{y})^2}}}
% \end{equation}

\section{Conclusions}
\label{conclude}
In this work we have developed a Combined Creativity score (CCS) to measure creativity in Scratch projects. CCS was developed based on an existing theory of creativity and focused on modeling the originality, elaboration and flexibility dimensions of a Scratch project based on the various elements of the project. 
We have run CCS on more than 400 Scratch projects taken from two studios, one for Maze game creation and the other for development of personal hero stories and games. 
We compared CCS to Dr. Scratch, an existing computational framework for measuring computational thinking in Scratch, and demonstrated the potential value of a creativity focused metric in automatically ranking projects based on creativity.
Feedback obtained from multiple Scratch teachers through interviews supported the theory used and the computational approach taken in this work. 
In future work we plan to correlate CCS with creativity scores given by expert Scratch instructors. We will also run CCS on additional Scratch Studios and plan to test its ability to estimate personal creativity progress, i.e. the evolvement of one student's creativity over time. 

%\section{Acknowledgments}

\bibliographystyle{authordate1}
\bibliography{main}

%\appendix

\balancecolumns

\end{document}